\documentstyle[12pt]{article}
\advance\textheight by \topskip
\def\edc{energy distinguished cores }
\begin{document}
\begin{flushright}
SU-ITP-94-25\\
hep-ph/9408247\\
July 1994
\end{flushright}
\vspace{.5 cm}
\thispagestyle{empty}
\begin{center}
{\Large
\bf ALIGNMENT IN GAMMA-HADRON FAMILIES\\
\vskip 1cm
OF COSMIC RAYS }

\par

\

\

\centerline {\bf V.V.Kopenkin$^1$, A.K.Managadze$^1$, I.V.Rakobolskaya$^{1,2}$,
T.M.Roganova$^1$}

\

$^1$ {\it Institute of Nuclear Physics, Moscow State University, Moscow 119899,
Russia}
\vskip 0.05cm
$^2$  {\it Department of Physics, Stanford University, Stanford
CA 94305, USA}
\end{center}

\vskip1cm
\par
\centerline {\large \bf Abstract}
\begin{quotation}

\rm     Alignment of main fluxes of energy in a target plane is found
in families of cosmic ray particles detected in deep  lead  X-ray
chambers.  The  fraction  of  events   with   alignment   is
unexpectedly large for families with high energy and  large  number
of hadrons. This can be considered as evidence for the existence
of coplanar scattering of secondary particles in  interaction  of
particles with superhigh energy, $E_0 {\
\lower-1.2pt\vbox{\hbox{\rlap{$>$}\lower5pt\vbox{\hbox{$\sim$}}}}\ } 10^{16}$
eV.

     Data analysis suggests  that   production  of  most  aligned
groups occurs low above the chamber and is characterized   by  a
coplanar scattering and quasiscaling spectrum  of  secondaries
in the fragmentation region.

     The most elaborated hypothesis for explanation of    alignment
is related to  the quark-gluon string rupture. However, the problem
of theoretical interpretation of our results still remains open.
\end{quotation}

\eject                                                        % -new page

\vskip14pt
\section { Introduction}

\rm International Pamir Collaboration is conducting a cosmic ray
experiment  at the altitude 4400 meters above  sea level in Pamir
mountains. Primary cosmic ray particles incident upon atmosphere produce
nuclear-electromagnetic cascades of secondaries in air. Hadrons and
electromagnetic particles related genealogically are called ``family".
Gamma-hadron family features depend on  interaction of hadrons with nuclei in
air.

Experimental data accumulated during more than 20 past years may allow us to
study interactions at very high energy (up to $E_0 \sim
10^{16}$). These energies are beyond the present accelerator range, and new
phenomena may reveal themselves in this region.

\vskip 20pt
 \subsection {Installation}

\rm Pamir Experiment equipment consists of X-ray emulsion chambers
of two kinds: carbon chambers (C-chambers) and deep lead chambers
 (Pb-chambers).

   Pb-chambers (see Fig.1) are assembled of many sheets
of lead of thickness 1 cm, interlaid with X-ray films. The total
depth of each Pb-chamber is from 40 cm ($\approx$70 c.u.) up to 110 cm
($\approx$195 c.u.). Thick lead substance provides both few interaction
lengths for hadrons and quasicalorimeter regime for energy
determination of particles.

   C-chamber \cite{[1]} consists of a block of 60 cm of
carbon covered on both sides by blocks of lead of thickness 6 cm at the
top and 5 cm at the bottom. Each block of lead contains 3 layers of
X-ray film. Carbon block provides the large cross section of hadron
interaction, while lead blocks are of minimal thickness allowing
determination of particle energies.

   Total area of chambers is few tens of square meters. Once a year all
these chambers are disassembled, the films are taken away and the
results of the experiment are investigated. The results to be reported in
this paper have been obtained by using deep Pb-chambers, which have some
advantages in hadron detection efficiency and energy determination
accuracy. On the other hand, C-chambers possess larger area of exposure.
Comparison
with some data from carbon chambers is  shown here too.

\vskip 20pt
 \subsection { Experimental procedure}

\rm Cosmic ray gamma-quanta and hadrons create electron-photon
cascades (or showers) in the lead.  (The term ``gamma-quantum"
is conventionally used for both gamma-quanta and electrons
(positrons)). These showers are detected in the
X-ray emulsion film as dark spots of a size which is typically
smaller than 1mm.

   The darkness density $D(E,t)$ of each spot depends on energy $E$
of the cascade and on the depth $t$ of its development in a
chamber. Comparing $D(E,t)$ for every shower with theoretical
predictions one can obtain the energy of each cascade and,
consequently, the energy $E_{\gamma}$ of a gamma-quantum incident upon
a chamber and producing this shower in it. By  doing  so for hadrons, one can
determine the energy $E_h^{(\gamma)}$ released into electromagnetic component
within the installation. It differs from the hadron energy
$E_h^0$ at the chamber surface by the factor $k_\gamma$ being around
${1/3}$ for pions.

   Gamma-quanta produce electron-photon cascades in the upper part of a
chamber only, whereas hadrons produce such showers at large depth as
well. The criterion for hadron identification in families is that the
breakthrough of a particle in a chamber (i.e. the shift of the origin of the
cascade
curve) has to be greater than 6 c.u.  In this case  only few
per cents of admixture of misidentified gamma quanta are present
among particles classified as hadrons.

   Efficiency of hadron detection is about 70-80 $\%$ in average for
Pb-chambers, and about 55 $\%$ for C-chambers.

   All chambers have energy determination threshold around 4 TeV for
$E_{\gamma}$ and $E_h^{(\gamma)}$ (or around 12 TeV for $E_h^0$
correspondingly).

   While dealing with gamma-hadron families one can reconstruct the
target diagram of an event by measurement of coordinates and incidence
directions of particles in the film emulsion. Thus one can find such
characteristics of a family as the total energy of gamma-quanta
$\Sigma$$E_{\gamma}$ or the total energy of hadrons  released to gamma-quanta
$\Sigma$$E_h^{(\gamma)}$, the distributions of gamma-quanta and
hadrons in the event area, $E_{\gamma}$ or $E_h^{(\gamma)}$ spectra,
etc. All families in our experiment were classified by the value of the
 total energy of the gamma-component $\Sigma$$E_{\gamma}$. Families with
$\Sigma$$E_{\gamma} \geq$ 100 TeV are under consideration here. When
studying ``superfamilies" with $\Sigma$$E_{\gamma} \geq$ 1000 TeV it
was found that in the central region of the event one can often see
one or few large diffuse dark spots (halos) in the X-ray films, of a size
from several millimeters up to few centimeters. Each such halo
appeared usually as a result of development of an atmospheric
electron-photon cascade from a high energy gamma-quantum produced at
some altitude above the chamber \cite{[2],[3]}.

   In the lower part of a deep lead chamber one can also find large spots
looking  like small halo, but having hadronic origin \cite{[4]}. Each
such halo is a result of a cascade produced by a hadron of a very high
energy in lead (with $E_h^{(\gamma)}$ about 200--500 TeV).
\vskip14pt

\vskip 20pt
 \subsection {History and formulation of the problem}

\rm     In 1985 Pamir Collaboration has found several families with
3 or 4 halos of electromagnetic origin \cite{[5],[6]},  and  in  most  of
these families (in 5 out of 6 such families) the multiple  halos
were aligned more or less along a  straight  line.  Experimental
results obtained during the subsequent years  did  not  increase
considerably statistics for investigation of  such  events,  but
the relative fraction of events with aligned multiple haloes  of
electromagnetic origin became smaller.

     As an alignment criterion  the  parameter  of  asymmetry
introduced by A.S.Borisov \cite{[7]} is conventionally used:
\begin{equation}\label{1}
{\lambda_m=}{ {\sum \limits _{i \not=j \not=k=1}^m
cos2\varphi_{ijk}} \over {m(m-1)(m-2)}} \  .
\end{equation}
Here $m$ is the number of objects, $i, j, k$ stand for vertices,
$\varphi_{ijk}$ is the angle between two vectors
$\overline{ki}$ and $\overline{kj}$.
An  event  is  considered  as aligned one if $\lambda \geq$ 0.6.
 (A  stronger requirement is $\lambda \geq$ 0.8).

    Parameter  $\lambda_m$ is the best known parameter of asymmetry
describing the degree of alignment rather than eccentricity.
For example, $\lambda_4$ will be equal 1 if four points belong to the
same straight line, but it will be far less than 1 if these points
form four vertices of a long rectangle.

     To  have  a  grasp  of  fluctuation  background,  i.e.  the
probability  of  random  occurrence  of  alignment   while   the
development  of  nuclear-electromagnetic  cascade,   a   computer
simulation of families with multiple  halos  was  made \cite{[8],[9]} using  a
quasiscaling model without any specific mechanisms for producing
asymmetry \cite{[10]}. Relative fraction of events with  three  aligned
halos  in  the simulated families appeared to be rather high,
about 30-35 $\%$ (by the criterion $\lambda_3 \geq$ 0.6).

     The  level  of  background  noise  calculated  for 3   random
incident  points (or ``particles" not belonging   to   the   same
cascade) was given by 24 $\%$ by  the  same  criterion. Therefore  the
appropriate analysis of the phenomenon to isolate the effect from
fluctuation background became   essential.

     However in the works \cite{[5],[6]} discussed above only halos at  the
same (small) depth in the upper  part  of  C-chambers  were
considered under some constraint on the level of darkness $D$ of the
spots  in X-ray films. Experimental results  obtained  in  Pb-chambers
allowed  to investigate alignment of multiple halos at different
observation depths and at various levels of darkness $D$, and  to take
into  account the contribution of hadron cascades (hadronic halos)
in the lower part of  a  chamber  \cite{[8]}. It  was  found  that  the
alignment of multiple halos in the same family is a  function  of
both  the depth  and  the  level  of  darkness  $D$ used for halo
identification. Therefore  this approach  seems   physically inadequate.

     It  is  worth  mentioning  here  that  an   attempt   to
investigate  asymmetry  of  family  particles configuration
(separately   for  gamma-component and for hadrons) in events
of small energies ($\Sigma E_{\gamma} =$ 100--400 TeV)  was made in
\cite{[11]}.
It was found that there was some excess of asymmetry in experimental events
over simulated ones. However, analysis was carried out with a quite
different criterion of asymmetry $\alpha$, and the existence of such
asymmetry did not necessarily imply alignment.

     In the investigation of alignment we tried to find a  better
method of selection of objects to be  examined,  which  would  be
more sensitive and less dependent on methodological factors. In
\cite{[9],[12]} it was suggested to  consider  not  only
halos, but a more general class of objects, which was called  ``Energy
Distinguish Cores" (EDC). These objects in  the  X-ray  film
correspond to the centers of the most prominent jets (air cascade
branches) with the highest energies in a family. They include the
following objects:

a) halos of the electromagnetic origin (or separate cores of
a multiple halo);

b) gamma-clusters  (i.e.  compact  groups  of  gamma-quanta
which  are  combined  into  clusters  using  the  criterion   of
decascading);

c) separate gamma-quanta of high energy;

d) high energy hadrons (in  particular,  the  hadrons  which
produced halos in the chamber).

   In order to treat gamma-component and hadrons in a  similar  way,
one should multiply by the factor of 3 the energy  $E_h^{(\gamma)}$
released by a hadron in the chamber into the electromagnetic
component, since the most secondaries in a family are pions and the average
fraction of energy transferred by pions to the electromagnetic component
 is approximately equal to $1 /3$.

     All energy distinguished cores (EDC)  are  considered  in
the order of decreasing energy, so it  becomes clear  how  to
select 3 or more objects in each family for analysis.

     This approach allows to study alignment in   gamma-hadron
families of not very high energies when there are  no  halos,  and  to
avoid discrimination of some types  of  EDC  against  some  other
ones.  By  this  method  the investigation   became   effective   and
physically  equivalent  for  both  charged  secondaries   in   an
atmospheric  shower  (family  hadrons)  and  neutral  secondaries
(gamma-component  in  the  same  family),   combining   them   to
 describe the interaction above a chamber.

     To investigate alignment of all family cores, which  are
detected at different depths in the chamber, a target diagram  was
made by projecting all traces of EDC onto one plane (for example,
onto the plane of the chamber top surface). If the zenith angle of
an event was not zero, the family image  was transformed to normal
plane. Alignment of the energy distinguished cores was studied
 in this plane, see e.g. Fig. 2.
\vskip14pt

\section{  Results}
\subsection{Experimental statistics}

\rm In this work we have analyzed  68 gamma-hadron families from deep
lead chambers with total energy of gamma-component $\Sigma E_{\gamma}
\geq $ 100 Tev, number of gamma-quanta $N_{\gamma} \geq$ 3 and number
of hadrons $N_h \geq$ 1 (see Table 1). In our data bank there are
also 19 families with $N_h=0$ which were not used in this work, since we were
looking only for gamma-hadron families.
Among these 68 events there are 18 families with
$\Sigma E_{\gamma} \geq $ 500 TeV, see Table 2. (Such high energy events were
collected
from larger area of installation than lower energy ones.)
 Total exposure of Pb-chambers here in use is about
450 m$^2\cdot$year. As for the hadron component, 13 events with $N_h >$ 10 are
present in our data.

     For comparison in some figures we also show the data from
carbon chambers of Pamir Russia-Japan Joint Experiment. This set
contains 84 gamma-hadron families with $\Sigma E_{\gamma} =
100-2600$ TeV (see Table 1) from total exposure around 440
m$^2\cdot$year. These results were obtained using Japanese X-ray
films from Pamir chambers measured in Waseda University (Tokyo)
and analyzed with participation of the authors.

\vskip 20pt
\subsection{ Evidence of alignment}

\rm     To find the effective criteria of alignment for analysis, we
tested various  threshold  values  of $\lambda$  and  variants  including
different numbers of cores (EDC) in a  family.  The  best  ratio  of  the
signal to the fluctuation background with satisfactory statistics appears with
the alignment criterion $\lambda_4 \geq$ 0.8. However, versions with other
numbers
of EDC are also  shown in our figures.

     The proposed approach allows  to  follow  behavior  of  the
fraction of events with alignment as a function of $\Sigma E_{\gamma}$,
  representing the family energy.

     In  Fig. 3(a,b,c)   one   can   see   such   dependence   for
correspondingly 3, 4 and 5 energy distinguished cores selected  in
each family in order of decreasing energy. Two  dashed  lines  in
each figure show levels of accidental occurrence of alignment in
model simulations (i.e. in artificial gamma-hadron  families)
and in simulated groups of randomly incident  objects.  One  can
see that fluctuation background level is always higher in model
events due to correlations in a cascade. The model which we have used
 \cite{[10]} does not involve any special mechanism of asymmetry.
Hereafter the criterion $\lambda \geq$ 0.8 is  used for
classification of the families with alignment.

   The increase of the fraction of events with alignment is evident for
families studied in deep lead chambers. This fraction rises from
the background level at $\Sigma E_{\gamma} =$ 100--300 TeV to $(61 \pm 18) \%$
for 3 cores under consideration and to $(47 \pm 17) \%$ for 4 EDC at
$\Sigma E_{\gamma} \geq $ 500 TeV.

   It is worth  noting that an additional analysis has been performed,   where
we have studied the behavior of an alignment fraction with energy when various
kinds of \edc
were considered separately. According to this analysis,
the fraction of events with alignment appeared to be independent
of energy for gamma-quanta under consideration, but increasing
with energy for both gamma-clusters and hadrons.   However, the
increase of alignment   with energy for gamma-clusters or
hadrons is less prominent than the similar rise for EDC, where these
two kinds of objects are included into consideration altogether with
halos and gamma-quanta.

    Such behavior of different kinds of \edc confirms indirectly
our understanding of the role of every component of a family in
the alignment phenomenon.

   The data obtained in the carbon chambers, Fig. 3, show the same
tendency as the data obtained in the lead ones, but  the increase of
alignment effect for C-chamber data   is somewhat less prominent.
There may be few reasons for this    difference.  The  data
   from C-chambers for maximal energy range are  poor in the high
energy events as compared with Pb-chambers. Besides that, the hadron
detection efficiency for C-chambers is considerably lower than for
Pb-chambers, and missed hadrons  may destroy the display of alignment
under consideration.

     In Fig. 3b one can see the estimated value of  the  fraction
of events with alignment if the thickness of lead chambers  were
be equivalent to the thickness of the carbon ones. This value seems to be in
agreement
with C-chambers data.  The accuracy of energy determination for
hadrons in carbon chambers (especially for high energy particles) is
also lower than in Pb-chambers, where the multilayer method allows to
follow a complete cascade curve from a particle in contrast to only one or two
points over a lead block in carbon chambers.

     The existence of the alignment effect    is  supported
by the fact that the experimental point at $\Sigma E_{\gamma} \geq$ 500 TeV
(see Fig.3b)  stays   at  two  standard deviations above     the  fluctuation
background level. If we estimate the combined  significance  of
deviation  from  the  background  for  2  independent  points  at
$\Sigma E_{\gamma} = $ 300--500 TeV and $\Sigma E_{\gamma} \geq$ 500
TeV, the $\chi ^2$-criterion   yields the confident level
$\simeq 99 \%$.

   Our model simulations of gamma-hadron families \cite{[13]}  showed
that the best correlation with the primary energy $E_0$ of the air cascade
is obtained not for  $\Sigma E_{\gamma}$  (or for
$\Sigma E_{total} = \Sigma E_{\gamma} + \Sigma E_h^{(\gamma)}$)  but  for
the number of hadrons in a family, $N_h$. Fluctuations of $N_h$
at fixed $E_0$ appeared to be 2 or 3 times less than fluctuations
of the total gamma-component energy  $\Sigma E_{\gamma}$ or
total hadron  energy  $\Sigma E_h^{(\gamma)}$  for  the same $E_0$.
Therefore, if the effect under consideration has energy threshold
while $E_0$ increases,  the  same  behavior  should  be observed
as a  function  of  hadron  number $N_h$,  the increase of alignment being
even
more distinct than while considering dependence on  $\Sigma E_{\gamma}$.

     In Fig. 4 the dependence of  the  fraction  of  events  with
alignment on the hadron number $N_h$ in a family is  presented  for
various   numbers   of   energy   distinguished    cores    under
consideration.   One can see evident rise of the  fraction  with
an increase of $N_h$. Thus the results presented in Fig. 4
confirm sensitivity of alignment to the number of hadrons in a
family.
 In Pb-chambers the fraction of families  with  alignment
comes to $(83 \pm 37) \%$ for 3 EDC and $(67 \pm 33) \%$ for 4 EDC.
The  increase of the effect for events from carbon  chambers  is
in  agreement with the Pb-chambers data. C-chamber families are poor
in events with large numbers of hadrons. Our comments on the carbon chambers
data shown in Fig. 3    are valid for this comparison too.

       Fig.5 shows the dependence of  the  fraction
of families with alignment on the number of energy  distinguished
cores in each family under investigation. One  can  see  that  in
families with  $N_h=1-3$  it  is  not  higher  than  the  level  of
fluctuation background, whereas for the group of  events  with
$N_h >$ 30 this fraction is much greater  than the calculated
background up to 7 cores considered. Despite  large  statistical
errors, this  makes  an  impressive  case  in  favour  of   reality   and
significance of the effect under consideration,  of its sufficiently high
frequency  of  occurrence  in  the  range   of   large  $N_h$   and,
consequently, of high energies of primary particle $E_0$.  The energy
scale  $E_0$  where we see a considerable  alignment begins at about $10^{16}$
eV.

      Fig. 5   also shows  the fraction of aligned events in  accelerator
data at $E_0 =$ 250 GeV  (target  experiment
NA22  at  CERN, $\pi$-Au  interaction \cite{[14]}).  These results  are in  a
remarkable
agreement with the results of our model simulations of the background level.
This confirms  the methods which we have used in our simulations, as well as
our conclusion that alignment is a threshold effect which occurs only at
sufficiently large energies.

\vskip 20pt
\subsection{ Transverse momenta of energy distinguished cores.}

\rm     The analysis of transverse momenta $p_t$  of EDC seems to be of
importance in theoretical explanation of the phenomenon. It is well
known that X-emulsion  chambers  detecting air families are able to
measure not $p_t$  itself, but a roughly  related quantity $ER$ (where $E$
is the particle energy, $R$ is the  distance  in the target plane
from an axis). Relation between $p_t$ and $ER$   is based  on
the assumption that the particles are produced in one interaction at some
altitude $H$ above an installation. In this case $p_tH=ER$.  $ER$  of
each  core  is  determined  in   reference   to   the
energy-weighed center of the group of 4 EDC in each family. Events
with  $\Sigma E_{\gamma} \geq$ 500 TeV were analysed. Average value
$<ER>$ in this  case appeared to be $2.1 \pm 0.8$ GeV$\cdot$ km for
events with alignment and $1.8 \pm 0.5$ GeV$\cdot$km for families
without alignment. One cannot see any significant difference in this
quantity between two classes of events.

     It seems  reasonable also  to  calculate  the  average  ratio  of
longitudinal $p_t^{\parallel}$  and  transverse $p_t^{\perp}$
(in  reference  to   the alignment direction in the target diagram plane)
in  the  same  events for the same 4 EDC in each. Such quantity
\begin{equation}\label{2}
{\Sigma p_t^{\parallel}}/{\Sigma p_t^{\perp}} =
                 {\Sigma ER^{\parallel}}/{\Sigma ER^{\perp}}
\end{equation}
is  similar to the famous parameter ``thrust". The average value
$<{\Sigma p_t^{\parallel}}/{\Sigma p_t^{\perp}}>$ was obtained to be
$\sim 11$ for events with alignment and $\sim 4$ for ones  without
alignment. This ratio differs considerably for the  two
cases. This is a natural consequence of separation
by the criterion $\lambda \geq$ 0.8.  Such evaluation for events
with alignment  enables us to see that  aligned cores come out
of complanarity plane by $<p_t^{\perp}> \sim 0.1<p_t>$.

     Thus assuming the most probable interaction altitude  $H =$ 2  km
(that follows from the halo superfamilies analysis \cite{[4]}), $<p_t>$  within
the group of  4  EDC  is  estimated  as  $\sim$ 1 GeV/c  and
 $<p_t^{\perp}> \approx 0.1 $ GeV/c.

\vskip 20pt
\subsection{ Energy distribution over the most energetic cores
                      in a family}

\rm     Energy distribution over 4 energy distinguished cores in each
family is another interesting characteristic.
     Fig. 6 shows the distributions in energy fraction
${E_i^{EDC}}/{\sum \limits_{i=1}^4{E_i^{EDC}}}$
for simulated families (quasiscaling MSF-model \cite{[10]}) in energy  ranges
$\Sigma E_{\gamma} =$ 100--500 TeV  and  $\Sigma E_{\gamma} >$ 500 TeV
and for  experimental  families  in  the same ranges. The shape of
the distribution does  not  change  with energy $\Sigma E_{\gamma}$
in  simulated  families,  and  the shape of the distribution
 for  low  energy  experimental
families agrees with simulations, whereas the plotted points for
superfamilies ($\Sigma E_{\gamma} >$ 500 TeV) considerably differ
from both above mentioned distributions.

     In this representation the  steeper  the function   falls  with
energy,  the  harder  the energy   spectrum   of   objects   under
consideration is. Solid line  shows  the  distribution  in  energy
fractions over the most energetic 4  particles  produced     in a direct
interaction in quasiscaling model at $E_0 = 10^{15}$ eV. It is evident
that the distribution for events with $\Sigma E_{\gamma} >$ 500  TeV
is  close  to  the calculated one for particles just after
an  interaction in the quasiscaling model.

     This shows that by investigation of the  energy  distinguished  cores  in
experiment we in fact study   the fragmentation  part  of  the
particle production spectrum, this part of the spectrum
being only slightly distorted by the air cascade and by the detecting device.

 In addition,  this   implies that the most energetic cores  in
the  majority  of  the  superfamilies  under  consideration   are
produced in one interaction at relatively low altitude above  the
chamber. (Particles coming  from a big altitude might  undergo  a
strong cascade affect).
\vskip14pt

\section{Discussion }

Alignment of energy distinguished cores  (or
particle streams,  or  energy  fluxes)  in  air  families
 should be related to a coplanar scattering  in  nuclear
interactions. It is very hard to explain the results of our experiment in the
framework  of  conventional interaction models.  It can   be inferred from
\cite{[15]} that the magnetic field of the Earth  could not be responsible for
any appreciable asymmetry. In the same work the obvious fact that  the coplanar
 ``fan"  of  particle  streams  may  be
blurred by cascade  process  after  few  interaction  paths  was
confirmed by model simulation. Therefore, either  the  interaction
which leads to the coplanar scattering occurs not far from the chamber, or it
may occur more than   few hadron interaction  lengths above the chamber.
However,  in  the last case  the
multiplicity of aligned particles in this ``fan" should be   large enough
to provide the   alignment  of     4  cores  at  the
observation level while other originally aligned particles drop out
of  the  original ``fan" plane due to the  cascade  development.

 There are two main  problems which should be solved in order to find a
theoretical explanation  of alignment. First of all, one should identify an
interaction mechanism, and then one should solve  the problem of intensity of
coplanar  events.  In the absence of a simple theoretical
interpretation of alignment, any  guess
on the possible interaction mechanism should be carefully considered.

     F.Halzen and  D.Morris   proposed  an  explanation  of
aligned multiple halos based on the    semi-hard  jet  model
($p_t > 3$ GeV)  \cite{[16]}. Such an interpretation does not seem   quite
satisfactory because
of  the difference in the energy distribution of  the main streams
(jet particles have too low energy).

     I.Roizen has suggested to interpret the phenomenon as a
projection of quark-gluon string rupture produced in the  process
of  semi-hard  double  inelastic  diffraction  dissociation,  the
string being inclined between a semi-hard  scattered  fast  quark
and  the  incident  hadron  remnants  \cite{[17]}.  Such   explanation   seems
plausible because the energy threshold of the alignment effect is
consistent with the threshold-like dependence  of  semi-hard  double
inelastic diffraction. The ``length" of aligned groups of EDC as a
projection is also more or less  in  agreement  with  the transferred
momentum while string production ($Q_t \simeq 3$ GeV/c). In this
case  the target diagram of a superfamily with alignment may be
considered as a direct ``photographic" image of such process.

     Average invariant mass M of the entire group  of  4  aligned
particles is $<M^2>=(60^{+120}_{-60}) GeV^2$. For the group of  6
aligned particles $<M^2>=(150 \pm 150) GeV^2$.  Such evaluation
of $M$ is again more or less compatible with inelastic diffraction
picture \cite{[17]}.

Note  that the energy range $E_0 \simeq 10^{15}-10^{16}$ eV is proclaimed as
the  threshold  for several unusual   processes: a) for  alignment  phenomenon;
b) for ``Centauro"  events production \cite{[18]}; c)  for  explanation of
electromagnetic particles spectrum in
extensive air showers in experiment ``Hadron" \cite{[19]} ; d) for semi-hard
double inelastic diffraction.

 Possible relation   of  alignment  to   the   string   rupture
hypothesis has been already mentioned above. It is  pertinent  to
add that the ratio ${<p_t^{\perp}>}\over {<p_t^{\parallel}>}$ $\sim 0.1$
within  EDC  group  is  roughly consistent with proper string
parameters in  transverse  momentum space, but as our preliminary
model simulations show, it  may  be appropriate to assume very small
$<p_t^{\perp}> \sim$ 20 MeV across the string in order to explain
of alignment. Such value has something in common with features
of  the hypothetical  ``Chiron"   events   production   \cite{[20]}
suggested to appear in the same energy range.

The authors understand that the ideas discussed above do not
constitute a complete  theoretical interpretation of alignment.
However, any hint can be
important when discussing events at such a
high energy and with such a  hard-to-reach  statistics.  Active  search  for
satisfactory explanation is necessary and is under way.

     It  would  be  most  desirable  to  test  this   effect   on
accelerators. Preliminary estimates indicate  that  the  energies
accessible at FNAL would be barely enough to  produce  comparable
families. However, one can still obtain interesting results at
these energies due to  the  possibility  of having  much  better
statistics than in cosmic rays.

\vskip 20pt

{\bf        Acknowledgements  }

     The  authors  would  like  to  express  their  gratitude  to
E.L.Feinberg,  I.L.Roizen,  S.A.Slavatinsky,   G.T.Zatsepin   for
active discussion of the results and to Y.Fujimoto and S.Hasegawa
for opening up opportunities to  use  the  data  bank  of  Waseda
University.   We   are   also   thankful    to    I.A.Mikhailova,
L.P.Nikolaeva,  E.G.Popova,  E.I.Pomelova,  L.G.Sveshnikova   and
N.G.Zelevinskaya for the large work they have done  participating
in Pb-chamber experiment. One
of the authors (I.R.) is greatly thankful to the Department of
Physics of Stanford University, and especially to R. Wagoner for
their kind hospitality  at Stanford, where this paper was completed.
     This  work   was      supported in part by International Scientific
Foundation.
\vfill
\newpage

\centerline{\bf
   Table 1}

\rm \parindent2cm
   Experimental events in use: the number of gamma-hadron families
with $\sum E_{\gamma} \geq$ 100 TeV, $N_{\gamma} \geq$ 3 and $N_h \geq$ 1.

\begin{center}

\vskip14pt
 \begin{tabular}{|l|c|c|c||c|}\hline
$\sum E_{\gamma} $  [TeV]&  100--300 & 300--500 & $>$ 500  & Total \\ \hline

Pb-chambers            & 35 &  15 & 18 &  68      \\ \hline
C-chambers (Rus.-Jap.) & 57 &  12 & 15 &  84      \\ \hline
\end{tabular}

   \end{center}

\vskip 1.5cm
\centerline{\bf
   Table 2}

\rm \parindent2cm
   Families from deep lead X-ray chambers with $\sum E_{\gamma} \geq$
500 TeV or $N_h >$ 10.

\begin{center}

\vskip14pt
 \begin{tabular}{|l|r|r|r|r|c|c|}\hline
Name of   &  $N_{\gamma}$  &  $\sum E_{\gamma} \quad $  &  $N_h$  &
 $\sum E_h^{(\gamma)} \quad$   &   Halo   &  Alignment  \\
$\quad$event  &  &  [TeV]$\quad$ &  & [TeV]$\quad$ &
                             & by criterion \\
& & {\small $E_{\gamma} \geq$ 4 TeV} & & {\small $E_h^{(\gamma)} \geq$ 4 TeV}
 &   & $\lambda_4 \geq 0.8$  \\ \hline
LoLiTa   & 386&    6140&      31&     699&       +   &  +   \\ \hline
Pb-45    & 312&    4574&      44&    1055&       +   &  +   \\ \hline
Pb-28    & 195&    3069&      59&     824&       +   &  +   \\ \hline
Pb-3703  & 180&    2559&      23&     690&       +   &  --- \\ \hline
Pb-53    & 120&    2071&      44&     727&       +   &  --- \\ \hline
Pb-8     & 192&    1964&      33&     621&       +   &  --- \\ \hline
Pb-6     &  91&    1521&      44&     816&       +   &  +   \\ \hline
Pb-54    & 111&    1291&      30&     336&       --- &  --- \\ \hline
F73-9    &  76&     949&      11&     297&       --- &  +   \\ \hline
Pb-20    &  61&     897&      22&     637&       +   &  +   \\ \hline
Pb-3704  &  47&     890&       7&     352&       +   &  --- \\ \hline
Pb-6012  &  48&     668&       4&      53&       +   &  +   \\ \hline
Pb-2     &  60&     752&       3&     130&       +   &  --- \\ \hline
Pb-2105  &  63&     687&       5&      56&       --- &  +   \\ \hline
Pb-6013  &  58&     794&      12&     188&       +   &  --- \\ \hline
Pb-58    &  75&     625&      23&    1086&       +   &  --- \\ \hline
Pb-4711  &  29&     575&       3&     144&       --- &  --- \\ \hline
Pb-5901  &  47&     501&       2&      71&        +  &  ---    \\ \hline
Pb-2201  &  35&     390&      12&     125&       --- &   +  \\ \hline
\end{tabular}

   \end{center}

\vfill
\newpage

\centerline{\Large  \bf    Figure Captions.}

\rm
\smallskip
{\bf Fig.1} \quad Structure of the most-used deep lead chamber of 60 cm
thickness
       from Pamir Experiment.

\smallskip
{\bf Fig.2} \quad An example of the target  diagram  with  energy
distinguished
cores for the event with alignment (the family Pb-6).  $\lambda_4 =$ 0.95.
     Figures in  the plot stand  for  energy  in  TeV
     (already multiplied by 3 for hadrons).

\parindent2cm  EDC: \qquad \
%{\Huge $\bigodot$}
 \circle{15}   is the halo of electromagnetic origin;

\parindent4cm  {\Large $\bullet$}\quad  is  the   hadronic halo;

               {$\oplus$\quad     are the high energy hadrons;}

               {\footnotesize $\bullet$}\quad are  the family gamma-quanta;

               {$+$\quad are other  hadrons of the family.  }

\smallskip
\parindent .7cm
{\bf Fig.3} \quad Dependence of the fraction of families  with  alignment  on
      total gamma-component energy of an event $\Sigma E_{\gamma}$.

\leftskip1cm   $N_{total}$ is the total number of families in a given energy
range;

     $N_{align}$ is the number of families with alignment in the same energy
range

      a) considering 3 energy distinguished cores (EDC)  in  each
          family;

       b) considering 4 EDC in each family;

       c) considering 5 EDC.

       Experiment: \quad {\vrule  width3mm height3mm depth0pt} \  \ is for
Pb-chamber data;  \quad  ${\,\lower0.9pt\vbox{\hrule \hbox{\vrule height 0.3 cm
\hskip 0.3 cm \vrule height 0.3 cm}\hrule}\,}$ \ \  is  for  C-chambers  of
the  Pamir   Joint  Experiment    (data    bank    of    Waseda
 University); \quad ${\,\lower0.9pt\vbox{\hrule \hbox{\vrule height 0.3 cm
\hskip 0.3 cm \vrule height 0.3 cm}\hrule}\,}$ \ \   is  for  the  estimate  of
 probable   result   for Pb-chambers having the reduced thickness equivalent
               to C-chambers one.

            Simulations: \quad -- -- -- -- -- \quad is for  the  artificial
families  by
                    quasiscaling   model   without   any   special
                    asymmetry. \quad
                 - - - - -\quad is for randomly incident objects.

\leftskip0cm
\smallskip
{\bf Fig.4}  \quad Dependence of the fraction of families  with  alignment  on
       the hadron number $N_h$  in a family.

       a), b), c) are for consideration of  3, 4  and  5  EDC  in  each
       family (see captions in Fig. 3).

\smallskip
{\bf Fig.5} \quad Dependence of the fraction of families  with  alignment  on
       the  number  of  energy  distinguished  cores  (EDC)  under
       consideration in each family.

\leftskip1cm      For $N_{total}$, $N_{align}$ see captions in Fig.3,

           - - - - - \quad  is for the model simulation.

          Experiment: \ \  {\vrule width9pt height9pt depth0pt}  \ \  is for
         the families   from  deep  lead  chambers with $N_h >$ 30.
    { $\bullet$}\ \  is for the families from  deep  lead  chambers
     with $N_h=1-3$.
           $\bigtriangleup$\ \  is for accelerator data at $E_0 =$ 250 GeV
           (experiment NA22 at CERN, $\pi$-$Au$ interaction).

\leftskip0cm
\smallskip

{\bf Fig.6} \quad  The  distribution  of  energy  fractions  over  the   most
        energetic 4 cores in a family.

\leftskip1cm
Experiment with deep lead chambers:

\parindent2cm   {\vrule width9pt height9pt depth0pt}\ \    is for families
                                  with $\Sigma E_{\gamma} > $ 500 TeV,

           $\bullet$\ \  is for those  with $\Sigma E_{\gamma} =$ 100--500 TeV.

\parindent1cm  Simulations by quasiscaling MSF-model [10].

\parindent2cm  - - - - - -\quad is for artificial families with any
                               $\Sigma E_{\gamma}$;

-----------\quad is for secondaries just in interaction
                              at $E_0 = 10^{15}$ eV.

\end{document}